\documentclass[11pt,a4paper]{article}
\usepackage{jcappub}

\def\be{\begin{equation}}
\def\ee{\end{equation}}
\def\beq{\begin{equation}}
\def\eeq{\end{equation}}
\def\bea{\begin{eqnarray}}
\def\eea{\end{eqnarray}}
\def\bml{\begin{subequations}}
\def\blea{\bml\begin{eqnarray}}
\def\elea{\end{eqnarray}\end{subequations}}

\usepackage{amsmath}

\newcommand{\Hub}{\mathcal{H}}
\newcommand{\B}{\mathbf}
\newcommand{\der}[2]{\frac{\partial#1}{\partial#2}}
\begin{document}

\title{Fluid Mechanics of Strings}
\author{Daniel Schubring and}

\emailAdd{schub071@d.umn.edu}

\author{Vitaly Vanchurin}

\emailAdd{vvanchur@umn.edu}

\date{\today}

\affiliation{Department of Physics, University of Minnesota, Duluth, Minnesota, 55812}

\abstract{

We consider conserved currents in an interacting network of one-dimensional objects (or strings). Singular currents localized on a single string are considered in general, and a formal procedure for coarse-graining over many strings is developed. This procedure is applied to strings described by the Nambu-Goto action such as cosmic strings. In addition to conserved currents corresponding to the energy-momentum tensor, we obtain conserved currents corresponding to an antisymmetric tensor $\langle F^{\mu\nu} \rangle = \langle x^{\prime\mu} \dot{x}^\nu - \dot{x}^\mu x^{\prime \nu} \rangle$, where $\dot{x}^\mu$ and $x^{\prime\mu}$  are the velocity and tangent vectors of strings. Under the assumption of local equilibrium we derive a complete set of hydrodynamic equations for strings.}

\maketitle

\section{Introduction}

A fluid description of zero-dimensional objects (or particles) can be derived from microscopic equations of motion by considering a coarse-grained evolution of distributions instead of individual particles. Then the microscopic conservation laws can be expressed as continuity equations of mass, momentum and energy that the distributions must obey. Using the kinetic theory of particles one can also show that there is an equilibrium distribution which is largely independent of the details of interactions. Then, under assumption of a local equilibrium, the conservation equations are reduced to a system of only five equations with five unknown parameters: (one) density, (three) velocity and (one) temperature fields. 

In this paper we will derive a fluid description of one-dimensional objects (or strings) whose microscopic evolution is governed by the Nambu-Goto action. Similarly to the particle fluid, the relevant effects of the microscopic interactions are captured by the kinetic theory whose central result is a derivation of the equilibrium distribution of strings \cite{Equilibrium}. Under the assumption of local equilibrium, we derive a complete system of seven equations which can describe the evolution of a string fluid regardless of the details of the interactions. The string fluid description is expected to be indispensable for the analysis of either topological or fundamental strings in the regimes where the conventional perturbative methods become unfeasible.

For example, networks of topological cosmic strings may form as the universe undergoes symmetry breaking phase transitions \cite{PhaseTransitions}. It is expected that such networks would give rise to very distinct and detectable signatures such as gravitational lensing \cite{Lensing}, CMB non-Gaussianities \cite{Nongaussianity}, gravitational waves \cite{Gravitywaves}, ultrahigh energy cosmic rays \cite{UHECR}, radio signals \cite{Radio} etc. (See \cite{Review} for a review of cosmic strings). It is also believed that cosmic superstrings could form at the end of brane inflation \cite{BraneInflation} which opens a possibility of testing string theory in the cosmological settings. Unfortunately, the networks of cosmic strings are usually analyzed using either numerical methods \cite{NewNumerical, OthersNumerical} within tight computational constraints or analytical models \cite{AnalyticKibble, NewAnalytic, AnalyticVanchurin} with limited ranges of validity. These limitations make it difficult, if not impossible, to obtain precise observational predictions that would be based on the statistical properties of cosmic strings. 

Another example are the networks of long fundamental strings which are expected to form above the Hagerdon temperature \cite{Hagerdon}. In addition to purely theoretical interests the high temperature effects may give rise to important observational signatures --- as argued, for example, by proponents of the string gas cosmology \cite{StringGas}. Indeed, small inhomogeneities in a network of fundamental strings could lead to primordial fluctuations as the universe cools down. Of course, such predictions would require an extensive analysis of fundamental strings in the Hagerdon phase. However, because of divergences in the canonical partition function, the usual methods of statistical mechanics are not very useful for calculating physical observables \cite{Hagerdon}. On the other hand, coarse-grained dynamics of either fundamental or topological strings could also be analyzed using the string fluid description developed in this paper, given that the local equilibrium does not depend on the details of the interactions.
 
This paper is organized as follows. In the Sec. \ref{Sec:Preliminaries} we discuss some basic properties of Nambu-Goto strings and in Sec. \ref{Sec:Currents} we derive continuity equations for singular currents of strings. In Sec. \ref{Sec:Fluid} we develop a coarse-grained description of strings and derive a set of hydrodynamic equations. The main results of the paper are discussed in Sec. \ref{Sec:Discussion}.

\section{Preliminaries}\label{Sec:Preliminaries}

The dynamics of a single string is well-described by the Nambu-Goto action, which can be expressed in terms of generalized worldsheet coordinates $ \zeta^a $,
\begin{equation}
S = -\int\sqrt{-h}\, d^2\zeta,
\label{action}
\end{equation}
where for simplicity the string tension is set equal to one. Here, $ h $ is the determinant of the metric on the world sheet, which is induced from the metric $ g_{\mu\nu} $ by pulling back the mapping into spacetime $ x^\mu(\zeta^a) $:
\begin{equation}
h_{ab} = g_{\mu\nu}\der{x^\mu}{\zeta^a}\der{x^\nu}{\zeta^b}.
\label{hmetric}
\end{equation}

By varying the action \eqref{action} with respect to $ g_{\mu\nu} $ we find the energy-momentum tensor $ T^{\mu\nu} $, 
\begin{equation} T^{\mu\nu}\sqrt{-g} = \int d^2\zeta \sqrt{-h}h^{ab}\der{x^\mu}{\zeta^a}\der{x^\nu}{\zeta^b}\delta^{(4)}\left (y^{\sigma}-x^\sigma\right ),
\label{4dT}
\end{equation}
where $ y^{\sigma} $ is the argument of $ T^{\mu\nu} $, and $ x^{\sigma} $ is again the mapping from the worldsheet into spacetime. (See \cite{Review} for details.) 

This expression \eqref{4dT} can be simplified by fixing our choice of $ \zeta^a $. The timelike coordinate will be denoted by $ \tau $ and the spacelike coordinate by $ \sigma $. We fix $ \tau $ to be equal to the spacetime coordinate $ x^0 $:
\begin{equation}
x^0(\tau,\sigma) = \tau.
\label{gauge1}
\end{equation}
Then the integration over $ \tau $ eliminates the temporal part of the delta function in the expression \eqref{4dT} for $ T^{\mu\nu} $:
\begin{equation}
T^{\mu\nu}\sqrt{-g} = \int d\sigma \tilde{T}^{\mu\nu}(\sigma)\delta^{(3)}\left(y^{i}-x^i\right ).
\label{current}
\end{equation}
The tilde notation $ \tilde{T}^{\mu\nu} $ indicates the non-singular part of the integrand. While $ T^{\mu\nu} $ is a singular density over spacetime, $ \tilde{T}^{\mu\nu} $ is a density over the worldsheet. This notation will be used for other tensor densities of the form \eqref{current} as well.

Denoting derivatives with respect to $ \tau $ and $ \sigma $ by dots and primes respectively, we adopt a further gauge condition on the worldsheet coordinates:
\be
\dot{\B{x}}\cdot\B{x}^\prime = 0.
\label{gauge2}
\ee
Restricting our consideration to the Friedmann universe in conformal coordinates with metric 
\be
 g_{\mu \nu}=a^2(\tau)\eta_{\mu \nu} ,\label{metric}
\ee
the energy density is given by
\be
\epsilon \equiv  \tilde{T}^{00}  = \sqrt{\frac{\B{x}^{\prime 2}}{1 - \dot{\B{x}}^2}}.
\label{epsilon}
\ee
If we also define the string velocity $ \B{v} \equiv \dot{\B{x}} $, the tangent vector $ \B{u} \equiv \epsilon^{-1}\B{x}^{\prime} $ and the Hubble parameter $ \Hub \equiv \dot{a}/a $, then the equation of motion is found to be\begin{equation}\dot{\B{v}} + 2\mathcal{H}(1-\B{v}^2)\B{v} = \epsilon^{-1} \B{u}^\prime\label{eqnFriedmann}.
\end{equation}
The quantity $ \B{v}^2+\B{u}^2 $ is a constant of motion which can be fixed by imposing a final gauge condition,
\begin{equation}
\B{v}^2 + \B{u}^2 = 1.
\label{gauge3}
\end{equation}

By applying these gauge conditions \eqref{gauge1}, \eqref{gauge2} and \eqref{gauge3} to equation \eqref{4dT} we can solve for the non-singular part of the energy-momentum tensor,
\begin{equation}
\tilde{T}^{\mu\nu} = \epsilon\,(v^\mu v^\nu -  u^ \mu u^ \nu).
\label{T}
\end{equation}
Here $ u $ and $ v $ have timelike components $ v^0 = 1 $ and $ u^0 = 0 $. So the energy density $ \tilde{T}^{00} = \epsilon $, the momentum density $ \tilde{T}^{i0} = \epsilon v^i $, and the spacelike components $ \tilde{T}^{ij} $ appear as the momentum current density in the continuity equation for momentum.

\section{Conserved Currents}\label{Sec:Currents}

\subsection{Minkowski Space}

In order to simplify the analysis of the energy-momentum tensor, we will first restrict our attention to Minkowski spacetime where $ \Hub = 0 $ and $ \epsilon = 1 $. In this special case, the equations of motion \eqref{eqnFriedmann} simplify to the wave equation,
\begin{equation}
\dot{\B{v}} = \B{u}^\prime.
\label{waveeq}
\end{equation}
Moreover, the conservation of the energy momentum tensor in flat spacetime can be expressed using an ordinary divergence, without additional gravitational correction terms, \begin{equation}
\partial_\mu T^{\mu \nu} = 0.
\label{diff_continuity}
\end{equation}
However, because of the delta functions in \eqref{current}, it is not immediately clear how to interpret the continuity (or conservation) equation \eqref{diff_continuity}. We will approach the problem by considering instead the integral form of the differential equation \eqref{diff_continuity} over an appropriate choice of enclosing volume. For a general current density $ j^\mu $ the integral equation is given by,
\begin{equation}
\partial_0 \int j^0 dV = - \int \B{j}\cdot d\B{A}.
\label{cont}
\end{equation}

\subsubsection{Particles}

When $ \B{j} $ is in the direction of the velocity $ \B{v} $, the situation is much the same as that of a localized particle. So we begin by considering the current density for a single particle,
\be
j^\mu = J^\mu \delta^{(3)}(y^{i}-x^i)
\ee
We choose a volume in \eqref{cont} which contains the particle for some time $ \tau<\tau_0 $. The particle leaves the volume at time $ \tau_0 $ and the boundary surface is chosen such that $ \B{v} $ is normal at the point where the particle exits.

By integrating \eqref{cont} over a small interval of time $ \Delta \tau $, the left hand side becomes the net change in enclosed charge, $ -J^0 $. We choose our coordinate system with $ x_\perp $ in the direction of $ \B{v} $, normal to surface. The current $ \B{J} $ is also in this normal direction, but in preparation for the more general case we will write $ \B{J}\cdot d\B{A} = J_v \, dA$. The integration over area in the flux integral cancels with the other two dimensions in the delta function and \eqref{cont} reduces to
\begin{align}-J^0 & = - \int\delta^{(3)}(y^{i}-x^i(\tau)) \, \B{J} \cdot d\B{A} \, d\tau \nonumber \\
	&= - \int J_v \, \delta^{(1)}(y_\perp-x_\perp(\tau))\,  d\tau \nonumber \\
	&= - \int J_v \, \delta^{(1)}(y_\perp-x_\perp) (\frac{dx_\perp}{d\tau})^{-1} dx_\perp \nonumber \\
	&= - J_v \, v^{-1}.
\end{align} 
Here the factor of $ v = dx_\perp / d\tau $ came about by changing our remaining integration variable to $ dx_\perp $. So the continuity equation for a localized particle just implies the familiar fact that the non-singular part of the current-density is the charge times the velocity,
\be
J_v = J^0 v.
\ee

\subsubsection{Strings}\label{sec:strings}

In the case of a string, in addition to the current in the direction of $ v^i $ it is physically relevant to have a current propagating along the string in the direction of $ u^i $. Even when a piece of string is contained in a volume it may pierce the surface at two or more points, and the flux of the current density at these points contributes extra terms in \eqref{cont}. 

Nevertheless, the argument for a localized particle can be extended straightforwardly to an infinitessimal piece of string with $ J^\mu = \tilde{J}^\mu d\sigma $. In the limit of $ \Delta \tau \rightarrow 0 $, only the terms due to the string discontinuously leaving the volume remain in the continuity equation. So the current in the direction of $ v^i $ follows the same expression as before,
\begin{equation}
\tilde{J}_v^i = \tilde{J}^0 v^i.
\label{jv}
\end{equation}

To consider the current in the direction of $ u^i $, we will first write an expression for the flux at a point where the string pierces the surface using the general form of the singular current. The coordinate $ x_\perp $ again points in the normal direction, and the $ \perp $ subscript denotes the $ x_\perp$-component of a vector. Then,
\begin{align}
\int \B{j}\cdot d\B{A} & = \int \delta^{(3)}(y^{i}-x^i(\sigma)) \, \tilde{\B{J}} \cdot d\B{A} \, d\sigma \nonumber \\
& = \int \delta^{(3)}(y^{i}-x^i(\sigma)) \, \tilde{J_\perp} dA \, d\sigma \nonumber \\
& = \int \delta^{(1)}(y_\perp -x_\perp(\sigma)) \, \tilde{J_\perp} \, d\sigma \nonumber \\
& = \pm \tilde{J_\perp} \, (\frac{dx_\perp}{d\sigma})^{-1} = \tilde{J_\perp} \, |x_\perp^\prime|^{-1}.
\label{flux}
\end{align}
Note that the change in variables leads to a negative sign if $ dx_\perp / d\sigma $ is negative, hence the use of the absolute value $|x_\perp^\prime| $.

The expression \eqref{flux} can be applied to the continuity equation for the momentum density $ T^{i 0} $ in \eqref{T} with $\epsilon=1$, where the associated current density has a term in the direction of $ u^k $, $ \tilde{J}_u^k = -u^i u^k $.  We choose a boundary surface surrounding a segment of string such that $ u^i $ is normal to the surface at the two points where the string enters and leaves the enclosed volume. The values of sigma at these points are denoted by $ \sigma_i $ and $ \sigma_f $, respectively. The left-hand side of the continuity equation \eqref{cont} becomes simply,
\be
\partial_0 \int T^{i0} dV = \partial_0 \int v^i \, \delta^{(3)}(y^{ i}-x^i) \, d\sigma \,dV = \int \der{v^i}{\tau} \,  d\sigma.
\ee
Since in flat spacetime $ \B{u} = \B{x}^\prime $, the component of $ \tilde{J}_u^k $ in the normal direction is just $ \tilde{J}_\perp = \mp u^i \, |x^\prime| $. So using \eqref{flux}, the continuity equation becomes,
\begin{align}
\int \der{v^i}{\tau} \,  d\sigma & = -(\left.\tilde{J}_\perp |x^\prime|^{-1}\right|_{\sigma_f}+\left.\tilde{J}_\perp |x^\prime|^{-1}\right|_{\sigma_i})\nonumber\\
& = u^i(\sigma_f) - u^i(\sigma_i)
\end{align}
which is just the equation of motion \eqref{waveeq} integrated over $ d\sigma $. Note that the equation of motion  \eqref{waveeq} has the form of a one-dimensional continuity equation on the worldsheet,
\begin{align}
 \der{\tilde{J}^0}{\tau} = -\der{\tilde{J}_\sigma}{\sigma}.
 \label{contsigma}
 \end{align}
In this case the charge density $ \tilde{J}^0 = v^i $ and the one-dimensional current density $ \tilde{J}_\sigma = -u^i $. 

In general, given any continuity equation of the form \eqref{contsigma} we can reverse the previous argument to find the singular current density in spacetime, $ \tilde{J}_u^i = \tilde{J}_\sigma x^{\prime i} $. This can be combined with \eqref{jv} for $ \tilde{J}_v^i $, to find the total current density,
\begin{align}
 \tilde{J}^k = \tilde{J}^0 \dot{x}^k + \tilde{J}_\sigma x^{\prime k}.
 \label{currentvu}
 \end{align}
 In particular, the commutation of partial derivatives is a continuity equation of the form \eqref{contsigma} 
 \be 
 \der{}{\tau}\left (\der{x^i}{\sigma}\right) = -\der{}{\sigma}\left (-\der{x^i}{\tau} \right ),
 \label{commute}
 \ee
 and by \eqref{currentvu}, this implies a conserved singular charge density $ \tilde{J}^0 = x^{\prime i} $ with an associated current density which we denote,
 \begin{equation}\tilde{F}^{ik} \equiv x^{\prime i} \dot{x}^k -\dot{x}^i x^{\prime k}.
 \label{Fcurrent}
 \end{equation}
 The conservation of this charge density depends only on the commutation of the partial derivatives of $ x(\tau,\sigma) $ and not on the Nambu-Goto dynamics.
 
 \subsubsection{Intersections}\label{sec:intersections}
 
 For each of the three components $ x^{\prime i} $, there is a continuity equation involving the flux of $ \tilde{F}^{ik}$. We may consider extending the expression \eqref{Fcurrent} to the timelike components,
 \be
 \tilde{F}^{0k} \equiv -x^{\prime k}, 
 \label{F0}
 \ee
 which motivates us to consider the fluxes of $x^{\prime k} $ as well.
 
 From the general expression for the flux of a singular current density \eqref{flux}, the flux of $ x^{\prime k}$ at a single intersection point equals,
 \be
 \tilde{J_\perp} \, |x_\perp^\prime|^{-1} = x_\perp^\prime \, |x_\perp^\prime|^{-1} = \pm 1.
 \ee
 The sign depends on whether $ x_\perp^\prime $ is parallel or antiparallel to the normal direction. Considering $ \B{x}^\prime $ to specify a direction of motion along the string, the sign depends on whether the string is leaving or entering the volume.
 
 In general, a string may intersect a closed surface at many points. As long as the string does not terminate in the interior (on a topological monopole for topological strings or on a $D$-brane for fundamental strings), for each point where the string enters the volume there must be another point at which the string leaves. So this means that the sum of the flux over all of these intersection points equals zero. Using \eqref{F0} we can express this as a flux integral of $ F^{0k} $ in space,
 \begin{equation}
 \oint F^{0k} dA_k = 0.
 \label{constraintint}
 \end{equation}
And so the top row of $ F^{\mu\nu} $ also obeys the continuity equation \eqref{cont}, with $ j^0 = F^{00} = 0$.

Just as a string can not terminate on a monopole in the interior, an intersection point on the surface can not suddenly disappear. An intersection point where a string leaves the volume can only vanish if it converges with a point where the string enters the volume. This suggests a picture in which the intersection points are two-dimensional particles with a charge of either $ \pm 1 $. A particle can only be created or annihlated in conjunction with an antiparticle of opposite charge. We wish to find the continuity equation for this flux charge. 

From our discussion on localized particles, it is clear that the corresponding current is just the charge multiplied by the two-dimensional velocity $ w^i $ on the surface. To find this velocity, we choose our coordinate system so that the surface near an intersection point is given by $ x^3 = 0 $. Similarly to \eqref{gauge1} which fixes our worldsheet coordinate $ \tau $, we introduce a new spatial worldsheet coordinate $ \zeta $ which is equal to $ x^3 $ in the vicinity of the intersection point. Formally, $ x^3(\tau^\prime,\zeta)= \zeta $ near the intersection point. For clarity, the transformed timelike coordinate is written as $ \tau^\prime $, even though $ \tau^\prime = \tau $. Then,
\begin{align}
w^k \equiv \der{x^k}{\tau^\prime} &=  \der{\tau}{\tau^\prime}\der{x^k}{\tau} + \der{\sigma}{\tau^\prime}\der{x^k}{\sigma} \nonumber\\
&= \dot{x}^k + \der{\sigma}{\tau^\prime} x^{\prime k} \label{weq}.
\end{align}
Since the partial derivative with respect to $ \tau^\prime $ is taken at fixed $ \zeta $, $ w^3 = 0 $. Thus,
\be
 \frac{\partial \sigma}{\partial \tau^\prime} = \frac{\dot{x}^3}{x^{\prime 3}},
 \label{w3eq}
\ee
and by substituting \eqref{w3eq} into \eqref{weq} we get
\be 
w^k = \dot{x}^k - \frac{\dot{x}^3}{x^{\prime 3}} x^{\prime k}.
\label{wkeq}
\ee

To find the two-dimensional singular current density we multiply this velocity by a two-dimensional delta function and the appropriate sign of charge. But since as before $ \pm 1 = \int x^{\prime 3} \, \delta(\zeta) \, d\sigma $, this can be `upgraded' to a three-dimensional string current density by multiplying $ w^k$ in \eqref{wkeq} by $ x^{\prime 3} $. So the charge density $ \tilde{J}^0 = x^{\prime 3} $ is conserved with current density 
\be
\tilde{J}^k= w^i  x^{\prime 3} =  x^{\prime 3} \dot{x}^k -\dot{x}^3 x^{\prime k}. 
\ee
But this is just the expression for the current density $\tilde{F}^{ik} $ in \eqref{Fcurrent}, only now the continuity equation involves flux through a surface rather than integration over a volume.

Abstracting back to the differential form of the continuity equation \eqref{diff_continuity}, it is easier to see how these two distinct integral continuity equations involving $ F^{ik} $ are related. Treating $ F^{i \nu} $ as a vector with index $ i $, we can consider the flux through a surface. Again choosing the coordinate system locally so that the normal is in the $ x^3 $ direction, $ \partial_0 F^{30} + \partial_k F^{3k}  = 0 $. But the current $ F^{3k} $ is clearly perpendicular to the normal $ k = 3 $ direction, so the current everywhere lies in the tangent space of the surface. So we can use a two-dimensional divergence theorem to bring \eqref{diff_continuity} into the form describing the conservation of intersection points discussed above.

\subsection{Friedmann Space}

So far we have been considering how densities on the worldsheet such as $ \tilde{T}^{\mu\nu} $ are related to singular densities in spacetime of the form
\be
\mathfrak{T}^{\mu\nu} \equiv  \int d\sigma \tilde{T}^{\mu\nu}\delta^{(3)}\left(y^{i}-x^i\right ).
\ee 
According to \eqref{current} the stress-energy tensor is related to $ \mathfrak{T}^{\mu\nu} $ through a factor of $ \sqrt{-g} $. In Friedmann space \eqref{metric} this factor $ \sqrt{-g} = a^4 $, and so
\be
T^{\mu\nu} = a^{-4}\, \mathfrak{T}^{\mu\nu}.
\label{tfrak}
\ee

In general relativity the continuity equation for the energy-momentum tensor involves the covariant divergence,
\be
0 = \nabla_\nu T^{\mu\nu} = \partial_\nu T^{\mu\nu} + \Gamma^\mu_{\lambda\nu}T^{\lambda\nu} + \Gamma^\nu_{\lambda\nu}T^{\mu\lambda}.
\label{covariant}
\ee
In Friedmann space the connection coefficients $ \Gamma^\mu_{\lambda\nu} $ all vanish except for
\be
\Gamma^0_{\mu\mu}=\Gamma^\mu_{0\mu}=\Gamma^\mu_{\mu 0}=\Hub.
\label{gamma}
\ee
So for any value of $ \nu $, $ \Gamma^\nu_{\lambda\nu} $ is nonzero only if $ \lambda = 0 $. Thus the last term in \eqref{covariant} reduces to
\be \Gamma^\nu_{\lambda\nu}T^{\mu\lambda} = 4\Hub T^{\mu 0} = 4\Hub a^{-4}\, \mathfrak{T}^{\mu 0} .\label{covariant_term3} \ee
Then by differentiating the first term in \eqref{covariant}, we find
\be
\partial_\nu T^{\mu\nu} = a^{-4}\, \partial_\nu \mathfrak{T}^{\mu\nu} - 4 \Hub a^{-4}\, \mathfrak{T}^{\mu 0},
\label{covariant_term1}
\ee
and the continuity equation \eqref{covariant} reduces to
\be
\partial_\nu \mathfrak{T}^{\mu\nu} + \Gamma^\mu_{\lambda\nu}\mathfrak{T}^{\lambda\nu} = 0.
\label{covariant_final}
\ee

Consider the momentum continuity equations, setting $ \mu = i $ and using \eqref{gamma}:
\begin{align}
0 &= \partial_\nu \mathfrak{T}^{i\nu} + \Gamma^i_{0i}\mathfrak{T}^{0i} + \Gamma^i_{i0}\mathfrak{T}^{i0}\nonumber\\
&= \partial_\nu \mathfrak{T}^{i\nu} + 2\Hub\mathfrak{T}^{i0}.
\label{covariant_momentum}
\end{align}
As before, this involves the time derivative of a charge density $ \tilde{J}^0 = \tilde{T}^{i0} $, and the divergence of a current density $ \tilde{J}^k = \tilde{T}^{ik}$. By \eqref{T},
\begin{align}
\tilde{T}^{ik} &= \epsilon\,(v^i v^k - u^i u^k)\nonumber\\
	&= (\epsilon\,v^i)\dot{x}^k + (-u^i)x^{\prime k},
\end{align}
so $ \tilde{J}^k $ takes the form of \eqref{currentvu}, leading to a continuity equation on the string. Here the only difference from \eqref{contsigma} is the gravitational correction term $ 2\Hub\tilde{T}^{i0} $ from \eqref{covariant_momentum}:
\begin{align}
0 & = \der{\tilde{J}^0}{\tau} + \der{\tilde{J}_\sigma}{\sigma} + 2\Hub\tilde{T}^{i0}\nonumber\\
& = \der{(\epsilon\,v^i)}{\tau} + \der{(-u^i)}{\sigma} + 2\Hub\epsilon v^i\nonumber\\
& = \epsilon\,\dot{v}^i + (\dot{\epsilon}+2\Hub\epsilon)v^i - u^{\prime i}.
\end{align}
So using the relation $ \dot{\epsilon}= -2\Hub \B{v}^2 \epsilon $ (see for instance \cite{Review}), we recover the equation of motion \eqref{eqnFriedmann} from a different perspective.

Unlike $ \tilde{T}^{\mu\nu} $, the conservation of $ \tilde{F}^{\mu \nu} $ depends only on topological properties (e.g. \eqref{commute}). So a conservation law of the form \eqref{cont} remains valid in Friedmann space without any gravitational correction terms. Still, $ \tilde{F}^{\mu \nu} $ in \eqref{Fcurrent} can be written in a form more appropriate to Friedmann space:
\begin{align}
\tilde{F}^{\mu \nu} &= x^{\prime \mu} \dot{x}^\nu -\dot{x}^\mu x^{\prime \nu}\nonumber\\
	&= \epsilon\,(u^\mu v^\nu - v^\mu u^\nu).
\label{F}
\end{align}

\section{String Fluid}\label{Sec:Fluid}

\subsection{Continuum Description}

As we have seen, the singular charge and current densities associated with a small segment $ \Delta \sigma $ of string with a given $ u $ and $ v $ take the form,
\be
q(x,u,v)=\tilde{Q}(u,v) \delta^{(3)}(x - y)\Delta \sigma 
\ee
where $ y $ is the position of the segment and $ x $ is the argument of the density function. We now consider a volume $ \Delta V $ containing many string segments as in Ref. \cite{Equilibrium}. The number of enclosed segments with parameters $ u $ and $ v $ is written as $ n(x,u,v)\Delta V $. Consider the integral of the charge density $ q $ over the coarse-graining volume $ \Delta V $. The delta function factor in $ q $ serves to count the number of enclosed segments and the integral becomes,
\be
\int \tilde{Q}(u,v) \Delta \sigma\, n(x,u,v)\,\Delta V\,du\,dv
\ee
Here $ \epsilon\Delta\sigma $ serves to convert the number density to an energy density, which is notated by $ f(x,u,v) \equiv \epsilon\Delta\sigma \, n(x,u,v) $. Dividing by the volume $ \Delta V $ we find the coarse-grained charge density,
\be
\langle \tilde{Q}\rangle \equiv   \int \epsilon^{-1}\tilde{Q} f(x,u,v) du\,dv.
\label{bracket}
\ee

Now consider the continuity equation \eqref{cont} involving the current density associated with $ \tilde{J}^\mu $. When the volume involved is much larger than $ \Delta V $, the average values $ \langle {J}^\mu\rangle $ may be used in the continuity equation. This approximation implicitly assumes that the distribution over $ u $ and $ v $ is statistically uniform at all points $ x_0 $ within the coarse-grained volume at $ x $. This can be abstracted to the case where $ \Delta V $ is infinitessimally small with respect to the volume of integration. Then the equation can be considered to be true for any volume, and we can pass to the differential form.

In particular, from \eqref{covariant_final} we obtain the following continuity equations,
\begin{equation}\partial_\nu \langle \tilde{T}^{\mu \nu}\rangle + \Gamma^\mu_{\lambda \nu}\langle\tilde{T}^{\lambda \nu}\rangle= 0
\label{Tcont}
\end{equation}
and
\begin{equation}
\partial_\nu \langle \tilde{F}^{\mu \nu}\rangle = 0,
\label{Fcont}
\end{equation}
where $ \tilde{T} $ and $ \tilde{F} $ are defined by \eqref{T} and \eqref{F} respectively.
Note that as in \eqref{covariant}, \eqref{Tcont} may instead be written as a covariant derivative of $ a^{-4}\langle \tilde{T}^{\mu \nu}\rangle $. Furthermore, since $ \langle \tilde{F}^{\mu \nu}\rangle $ is antisymmetric, \eqref{Fcont} may also be written in terms of a covariant derivative, $ \nabla_\nu \langle \tilde{F}^{\mu \nu}\rangle = 0 $.

Evaluating the connection coefficients in \eqref{Tcont} explicitly using \eqref{gamma}, we find the energy continuity equation:
\begin{align}
\partial_\nu \langle \tilde{T}^{0\nu}\rangle &=  -\Hub\sum_\lambda\langle\tilde{T}^{\lambda\lambda}\rangle\nonumber\\
&= -\Hub\langle\epsilon(1+(\B{v}^2-\B{u}^2))\rangle\nonumber\\
&= -2\Hub\langle\epsilon\B{v}^2\rangle,
\label{Tcont_energy}
\end{align}
and following \eqref{covariant_momentum} we have the momentum continuity equation:
\be
\partial_\nu \langle \tilde{T}^{i\nu}\rangle = -2\Hub\langle\epsilon v^i\rangle.
\label{Tcont_momentum}
\ee
Also note that the top row of \eqref{Fcont} does not involve a time derivative, and expresses the differential form of \eqref{constraintint}:
\begin{equation}
 \partial_i \langle \epsilon u^i \rangle = 0.
 \label{constraint}
\end{equation}

\label{ab}

The continuity equations \eqref{Tcont} and \eqref{Fcont} express the time derivatives of the fields $  \langle{\epsilon v^i}\rangle $ and $ \langle{\epsilon u^i}\rangle $ in terms of spatial derivatives of correlations such as $ \langle{\epsilon u^i u^j}\rangle $.  Rather than taking a thermodynamic approach at this point \cite{Blackfold}, we will simplify the equations under the condition of local equilibrium. To express this condition, it is helpful to consider a slightly different set of fields.

A solution to the equation of motion in flat space \eqref{waveeq} for a single string can be expressed in terms of two waves moving in opposite directions
\begin{equation}
x^i(\tau,\sigma) = \frac{a^i(\sigma - \tau) + b^i(\sigma + \tau)}{2}.
\label{ab_flat}
\end{equation}
Then it is convenient to consider the quantities $ A^i \equiv \partial a^i / \partial \tau  $ and $ B^i \equiv \partial b^i / \partial \tau  $ which can be expressed in terms of $ u $ and $ v $:
\begin{align}
 A^i &= v^i - u^i, \\
  B^i &= v^i + u^i.
\label{ab_definition}
\end{align}
The gauge condition \eqref{gauge3} implies both $ \B{A} $ and $ \B{B} $ are unit three-vectors. We can also extend the definitions of $ A^i $ and $ B^i $ to four-vectors with a timelike component of $+ 1$. 

Although \eqref{ab_flat} does not hold in Friedmann space, we can still define $ \B{A} $ and $ \B{B} $ using \eqref{ab_definition}. By \eqref{ab_flat}, in Minkowski space $ \B{A} $ and $ \B{B} $ are constant on paths of constant phase $ \sigma \mp \tau $. Likewise, in Friedmann space the dynamics of $ \B{A} $ and $ \B{B} $ greatly simplifies along certain paths on the world sheet \cite{BB}. Explicitly, the two families of paths $ (\tau(t_\pm),\sigma(t_\pm)) $ can be defined by,
\begin{align}
\frac{d\tau}{dt_\pm} &= 1\nonumber\\
\frac{d\sigma}{dt_\pm} &= \pm \epsilon^{-1}.
\label{ab_path}
\end{align}
Then using the equation of motion \eqref{eqnFriedmann}, the time derivatives of $ \B{A} $ and $ \B{B} $ simplify along these paths:
\begin{align}
\frac{d\B{A}}{dt_+}&=-\Hub(\B{B}-(\B{A}\cdot\B{B})\B{A})\\
\frac{d\B{B}}{dt_-}&=-\Hub(\B{A}-(\B{A}\cdot\B{B})\B{B}).
\label{ab_derivative}
\end{align}
So the quantity $ \B{A} $ might be thought of as moving along the paths parametrized by $ t_+ $, and $ \B{B} $ along those parametrized by $ t_- $. Their spatial velocities are then,
\begin{align}
 \frac{d\B{x}}{dt_+}&= \B{B}\\
\frac{d\B{x}}{dt_-}&= \B{A}.
\label{ab_velocity}
\end{align}
So in this picture $ \B{A} $ can be thought of as moving with velocity $ \B{B} $, and vice-versa.

Note that the symmetric and antisymmetric parts of the tensor product $ B\otimes A $ are just $ \tilde{T}^{\mu\nu} $ and $ \tilde{F}^{\mu\nu} $, respectively,
\begin{align}
 \langle \tilde{T}^{\mu\nu}\rangle = \langle \epsilon B^{(\mu} A^{\nu)}\rangle,  \\
\langle \tilde{F}^{\mu\nu}\rangle = \langle \epsilon B^{[\mu} A^{\nu]}\rangle.
\label{asymm}
\end{align}
Then we can rewrite the continuity equations \eqref{Tcont} and \eqref{Fcont} in terms of $ A $ and $ B $ fields as
\begin{equation}
\der{ }{\tau}\langle{\epsilon A^i}\rangle + \der{}{x^j} \langle{\epsilon A^i B^j}\rangle  = -\Hub\langle\epsilon (A^i + B^i)\rangle
\end{equation}
\begin{equation}
\der{ }{\tau}\langle{\epsilon B^i}\rangle + \der{}{x^j} \langle{\epsilon B^i A^j}\rangle = -\Hub\langle\epsilon (A^i + B^i)\rangle
\end{equation}

\subsection{Local Equilibrium}

We can use the new variables in the argument of the energy-density $ f(x,A,B) $. The energy-density function involves many small segments of strings in a given coarse-grained region of space and these segments may interact through reconnections or (if they happen to lie on the same string) through the Nambu-Goto dynamics  \cite{Equilibrium}. By modeling these interactions as an exchange of $ A $ and $ B $ vectors, a transport equation for $ f(x,A,B) $ may be derived. If $ f(A,B) $ is homogenous in space, it has been shown  \cite{Equilibrium} that an equilibrium distribution $ \partial f_{eq} /\partial \tau = 0 $ may be factored into parts depending only on $ A $ and $ B $ separately,
\begin{equation}
f_{eq}(A,B) \sim f_A(A) \, f_B(B).
\label{separable}
\end{equation}

We can treat $ f $ as probability distribution, defining the normalized expection value in terms of the coarse-graining brackets \eqref{bracket},
\be 
\bar{Q} \equiv \rho^{-1} \langle{\epsilon Q}\rangle, 
\ee
where the energy density $ \rho $ is the normalization factor,
\be
\rho \equiv \int f(A,B) dA\, dB.
\ee 
Then \eqref{separable} implies that at equillibrium $ A^i $ and $ B^j $ are independent random variables:
\begin{equation}
\langle{\epsilon A^i B^j}\rangle =  \rho \, {{\bar{A}}^i} {\bar{B}^j}.
\label{localeq}
\end{equation}
In the general case where $ f $ varies in space, we will likewise take `local equillibrium' to mean that $ A^i $ and $ B^j $ are independent at each point of space. 

On the other hand, $ u^i $ and $ v^j $ are not in general independent, but using \eqref{asymm} we can still factor both $ T^{\mu\nu} $ and $ F^{\mu\nu} $ into $ \bar{u}^i $ and $ \bar{v}^i $:
\be
\langle T^{\mu \nu}\rangle = \rho ( \bar{v}^\mu \bar{v}^\nu -\bar{u}^\mu \bar{u}^\nu),
\ee
\be
\langle F^{\mu \nu}\rangle = \rho ( \bar{u}^\mu \bar{v}^\nu -\bar{v}^\mu \bar{u}^\nu).
\ee

Because $ \B{A} $ and $ \B{B} $ are unit vectors, the variance does not depend on higher order moments:
 \begin{align}
 \text{Var}(\B{A}) &= \overline{\B{A}^2} - \bar{\B{A}}^2\nonumber\\
  &= 1 - \bar{\B{A}}^2\nonumber\\
  \text{Var}(\B{B}) &= 1 - \bar{\B{B}}^2.
  \label{var_AB}
 \end{align}
 And since $ \B{u} $ and $ \B{v} $ are linear combinations of the independent $ \B{A} $ and $ \B{B} $,
  \be
  \text{Var}(\B{v}) = \frac{1}{4}(\text{Var}(\B{A}) + \text{Var}(\B{B})) = \text{Var}(\B{u}).
  \label{var_independent}
  \ee
This can be expressed solely in terms of $ \B{u} $ and $ \B{v} $ using the gauge condition \eqref{gauge2}:
 \begin{align}
 \text{Var}(\B{u}) =  \text{Var}(\B{v}) &= \frac{1}{4}(2 - (\bar{\B{A}}^2+\bar{\B{B}}^2))\nonumber\\
 &= \frac{1}{2}(1 - (\bar{\B{u}}^2+\bar{\B{v}}^2)).
 \label{var_uv}
 \end{align}
 
 So the variance of $ \B{u} $ and $ \B{v} $ is related to the extent to which the gauge condition \eqref{gauge3} is violated by the averaged fields. Likewise, the condition \eqref{gauge2} is violated whenever $ \text{Var}(\B{A}) \neq \text{Var}(\B{B}) $. Using \eqref{var_AB}, it is easy to show,
 \be
 \text{Var}(\B{A}) - \text{Var}(\B{B}) = \frac{1}{4}(\bar{\B{v}}\cdot\bar{\B{u}}).
 \ee
 
 These expressions involving second order moments are useful in dealing with the factor of $ \langle\epsilon\B{v}^2\rangle $ in the gravitational correction to the energy continuity equation \eqref{Tcont_energy}. From \eqref{var_uv},
 \be
 \overline{\B{v}^2} = \frac{1}{2}(1 + (\bar{\B{v}}^2-\bar{\B{u}}^2)).
 \label{v2}
 \ee
 
\subsection{Fluid Equations}

The continuity equations can now be put in the familiar form of fluid mechanics. Ignoring the gravitational terms for now, we can write \eqref{Tcont} as the two equations,
\begin{equation}
\der{\rho}{\tau} + \der{}{x^j}(\rho \bar{v}^j) = 0
\label{rhoCont}
\end{equation}
and
\begin{equation}
\rho\left (\der{\bar{v}^i}{\tau} + \bar{v}^j \der{\bar{v}^i}{x^j} \right ) = \der{\sigma^{ij}}{x^j}.
\label{sigmaij}
\end{equation}
where the Cauchy stress tensor is defined as $ \sigma^{ij} \equiv \rho \, \bar{u}^i \bar{u}^j $. The stress tensor can be decomposed into a scalar `pressure',
\be
p \equiv -\frac{1}{3} \text{Tr}(\sigma)
\ee
and a traceless `viscous stress tensor' 
\be
\varepsilon^{ij} \equiv \sigma^{ij} + p\,\delta^{ij}.
\ee
With these definitions we can put \eqref{sigmaij} into the general form of the Navier-Stokes equations,
\begin{equation}
\rho \frac{D\bar{v}^i}{D\tau} = -\frac{\partial p}{\partial x^i} + \frac{\partial \varepsilon^{ij}}{\partial x^j} 
\label{navier}
\end{equation}
where the material derivative 
\be
\frac{D}{D\tau} \equiv \frac{\partial}{\partial \tau} + \B{\bar{v}} \cdot \nabla.
\ee
We stress, however, that \eqref{navier} differ from the proper Navier-Stokes equations in that the viscous stress tensor $\varepsilon^{ij}$ can not be written in terms of spatial derivatives of $ v $ times a viscosity coefficient.

Although $ p $ formally acts like the pressure, it is not clear whether it can be identified with the thermodynamic pressure. If there is a distinction, the viscous stress tensor may be defined with a nonzero trace in which case there would be a non-vanishing bulk viscosity \cite{Landau}. Also note that the energy-momentum tensor $ \rho ( \bar{v}^\mu \bar{v}^\nu -\bar{u}^\mu \bar{u}^\nu) $ is not in the form of a perfect fluid. But the condition that $ \varepsilon^{ij} $ vanishes implies that $ -\rho \, \bar{u}^i \bar{u}^j = p \delta^{ij}$. This condition is just what is needed to put the energy-momentum tensor in the form of a perfect fluid with pressure $ p$. So $ p $ is consistent with the pressure as defined in familiar cosmological models.

In general, it is a lot more informative to rewrite the hydrodynamic equations with a dynamical vector field $ \B{u} $ rather than the pressure and viscous tensor.  Using \eqref{v2} to simplify $ \overline{\B{v}^2} $ in the energy continuity equation \eqref{Tcont} we find,
\be
\der{\rho}{\tau} + \nabla \cdot(\rho \B{\bar{v}}) = -\Hub (\B{\bar{v}}^2-\B{\bar{u}}^2 + 1)\,\rho,
\label{eqrho}
\ee
and again from \eqref{constraint},
\be
\nabla \cdot(\rho \B{\bar{u}}) = 0.
\label{eqconnect}
\ee
Using these two equations to simplify \eqref{Tcont_momentum} and \eqref{Fcont},  we find,

\be
\frac{D\B{\bar{v}}}{D\tau} - (\B{\bar{u}}\cdot\nabla ) \B{\bar{u}} = \Hub (\B{\bar{v}}^2-\B{\bar{u}}^2 -1)\,\B{\bar{v}}
\label{eqv}
\ee
and
\be
\frac{D\B{\bar{u}}}{D\tau} - (\B{\bar{u}}\cdot\nabla ) \B{\bar{v}} = \Hub(\B{\bar{v}}^2-\B{\bar{u}}^2 +1)\,\B{\bar{u}}
\label{equ}
\ee
Note that the evolution of the $ \B{\bar{u}} $ and $ \B{\bar{v}} $ fields decouple from the energy density $ \rho $. 

We can also rewrite the decoupled equations \eqref{equ} and \eqref{eqv} in terms of the $\B{\bar{A}}$ and $\B{\bar{B}}$ fields using \eqref{ab_definition}, 
\be
\frac{\partial \B{\bar{A}}}{\partial \tau} + (\B{\bar{B}}\cdot\nabla) \B{\bar{A}} = -\Hub(\B{\bar{B}}-(\B{\bar{A}}\cdot\B{\bar{B}})\B{\bar{A}}),
\label{eqA}
\ee
\be
\frac{\partial \B{\bar{B}}}{\partial \tau} + (\B{\bar{A}}\cdot\nabla) \B{\bar{B}} = -\Hub(\B{\bar{A}}-(\B{\bar{A}}\cdot\B{\bar{B}})\B{\bar{B}}).
\label{eqB}
\ee
As discussed in relation to \eqref{ab_velocity}, $ \B{\bar{A}} $ can be considered to move with velocity $ \B{\bar{B}} $ and vice-versa. In this respect, the left hand sides of equations  \eqref{eqA} and \eqref{eqB} can be interpreted as material derivatives. So the material derivatives of the fields $ \B{\bar{A}} $ and $ \B{\bar{B}} $ are formally identical to the path derivatives \eqref{ab_derivative} for a single string. This is an intuitive, but non-trivial result given that the quantities appearing in \eqref{eqA} and \eqref{eqB} are the local averages of the $ \B{A} $ and $ \B{B} $ values over many string segments. In fact there is no reason to expect that the same equations would describe more general fluids in which the local equilibrium assumption is violated.

\section{Discussion}\label{Sec:Discussion}

We shall now discuss some of the immediate consequences of the sting fluid described by equations \eqref{eqrho}, \eqref{eqconnect}, \eqref{equ}, \eqref{eqv}. In the limit where the string fluid consist of only closed loops with typical sizes smaller than the coarse-graining scale, the average value of the tangent vector must vanish (i.e. ${\bf \bar{u}}=0$). As a result \eqref{eqv} in Minkowski space reduces to the inviscid Burgers' equation
\be
\frac{\partial  \B{\bar{v}}}{\partial \tau} + \left ( \B{\bar{v}} \cdot \nabla\right )  \B{\bar{v}} = 0,
\label{Burgers}
\ee
whose solutions are know to develop discontinuities (or shock waves) that can only be resolved with higher order terms. The small loops phase is relevant for describing cosmic strings at late cosmological times or fundamental strings below Hagerdon temperature and it would be interesting to study the observable signatures of such shock waves. 

More generally the string fluid might have a non-vanishing component of long strings (i.e. ${\bf \bar{u}}\neq0$) in which case the decoupled equations \eqref{eqA} and \eqref{eqB} must be solved first. This would be relevant for the analysis of the network of fundamental strings above Hagerdon temperature or cosmic strings in the early universe. In particular, one might be interested in the production and subsequent evolution of closed loops in a network of cosmic strings. In the language of sting fluids such processes would correspond to a monotonic decay of the $\bf{\bar{u}}$ field. Then it should be possible, for example, to distinguish the decay of infinite strings into small loops with typically large velocities ${\bf \bar{v}} \sim 1$ from the decay of infinite strings into large loops with typically small velocities ${\bf \bar{v}} \ll 1$.

Another important result of the string fluid discussion which is worth emphasizing again is the decoupling of equations \eqref{equ} and \eqref{eqv} from the other equations \eqref{eqrho} and \eqref{eqconnect}. This means that one can first solve for $\bf{\bar{u}}$ and  $\bf{\bar{v}}$ fields regardless of the energy density $\rho$ given that \eqref{eqconnect} is satisfied at some moment of time. Then \eqref{eqconnect} will be automatically satisfied at all times for $\rho$ which solves \eqref{eqrho}. Unfortunately, this also means that the obtained equations cannot describe the Hagerdon phase where the long strings are expected to form (i.e. ${\bf \bar{u}} \neq 0$) only when $\rho$ is sufficiently large. This suggests that the higher order non-equilibrium effects must be included to describe the fundamental strings at very high energy densities. 

In conclusion, we note that even without the local equilibrium assumption \eqref{separable} the continuity equations \eqref{Fcont} are of the same form as the homogenous Maxwell equations. In Minkowski space the equation \eqref{constraint} corresponding to $\mu=0$ in \eqref{Fcont} is analagous to the statement that there are no `magnetic' monopoles, 
\be
\nabla \cdot \langle \B{u}\rangle =0,
\ee
and the other rows of \eqref{Fcont}  corresponding to $\mu=1,2,3$ can be written in a way analogous to Faraday's law:
\begin{equation}
\der{\langle \B{u}\rangle}{\tau} = - \nabla \times \langle \B{u \times v}\rangle.
\end{equation}
In this perspective, the time derivative of the flux of a `magnetic' field $ \langle\B{u}\rangle $ is related to the circulation of an `electric' field $ \langle \B{u \times v}\rangle$, whereas before we were considering the flow of a two-dimensional `flux-current' across a one-dimensional boundary. Of course, the two pictures are mathematically equivalent, and it remains to be seen whether the field picture is useful.

\end{document}